\documentstyle[12pt]{article}

{   % make it in effect now, not after \begin{document}
\baselineskip\baselinestretch\baselineskip
\newcount\valueone  \valueone=\textheight
\newcount\valuetwo  \valuetwo=\baselineskip
\divide\valueone by \valuetwo
\advance\valueone by -2 % take into account two lines for footer
\message{number of lines per page = \the\valueone}
\global\textheight\valueone\baselineskip}
\def\erf{\mathop{\rm erf}\nolimits}
\def\Ln{\mathop{\rm Ln}\nolimits}
\def\Re{\mathop{\rm Re}\nolimits}

\begin{document}

\begin{titlepage}
\vspace{2cm}

\hfill {Preprint IHEP 95--70}
\vspace{3cm}

\centerline {\large V.~A.~Slobodenyuk }
\vspace{2cm}

\centerline{\bf \Large Analytical Properties of Solutions of
the Schr\"odinger Equation }
\centerline {\bf \Large and Quantization of Charge }
\vspace{2cm}

Moscow State University Branch, 432700 Ulyanovsk, Russia.

E-mail: slob@ftf.univ.simbirsk.su
\vspace{2cm}

\begin{abstract}
The Schwinger--DeWitt expansion for the evolution operator kernel is
used to investigate analytical properties of the
Schr\"odinger equation solution in time variable. It is shown, that
this expansion, which is in general asymptotic, converges
for a number of potentials (widely used, in particular, in
one-dimensional many-body problems), and besides, the convergence
takes place
only for definite discrete values of the coupling constant. For other
values of charge the divergent expansion determines the functions
having
essential singularity at origin (beyond usual $\delta$-function). This
does not permit one to fulfil the initial condition. So, the function
obtained
from the Schr\"odinger equation cannot be the evolution operator
kernel.
The latter, rigorously speaking, does not exist in this case. Thus,
the kernel exists only for definite potentials, and moreover, at the
considered examples the charge may have only quantized values.
\end{abstract}

\end{titlepage}

\sloppy

\section{Introduction}

In the quantum theory expansions in different parameters such as
the coupling constant~\cite{BW1,BW2,BW3,Lip}, the WKB--expansion, the
short-time Schwinger--DeWitt expansion~\cite{Sch,DeW1,DeW2,S2}, the
perturbation expansion in phase-space technique~\cite{BOG},
$1/n$-expansion~\cite{P}, etc. are, as a rule, asymptotic.
This circumstance imposes essential restrictions on possibilities
of their using, makes the theory incomplete and compels one to look
for the ways of overcoming these restrictions
either by summation of divergent series with special methods (see,
e.g.,~\cite{KS}), or by constructing new convergent
expansions~\cite{HS,Ush,SS},
 or by creating different approximate methods taking into
consideration the so-called nonperturbative effects.

Useful information may be obtained from investigation of the character
of singularity causing the divergence of expansions, as it was made,
e.g., in~\cite{BW1} for the expansion in the coupling constant for the
anharmonic oscillator. The purpose of the present paper is to
investigate
analytical properties of the evolution operator kernel of the
Schr\"odinger
equation in time variable in the neighborhood of origin. As usual,
one works
with the short-time Schwinger--DeWitt expansion~\cite{Sch,DeW1},
having numerous applications mainly connected with the theories in the
curved space--time~\cite{DeW2,BV}, as with asymptotic one. However,
it can be helpful for examining analytical properties of the kernel in
vicinity of the point $t=0$, in particular, for searching for the
potentials,
for which this expansion is convergent and the point $t=0$ is the
regular one.

An important feature of the Schwinger--DeWitt expansion is that after
factorization of the contribution of the free kernel (``free'' case
corresponds to $V\equiv 0$),
having at $t=0$ the singularity in the form of $\delta$-function,
one can concentrate attention on rest part which, according to the
initial condition, should be equal to 1 when $t=0$. The essential
point of presented research is: is it possible or not to satisfy this
initial condition meaning $t$ as complex variable?

If one considers the real $t$ only (this is usually done), then
analytical
properties in the vicinity of the point $t=0$ are fully ignored.
If the
analytical continuation into the complex plain $t$ is made for the
kernel
(this may be done only inside some sector with angle $\gamma < \pi$),
then
its analytical properties are masked by the singularity in the form of
$\delta$-function. But if the factorization of the free part of the
kernel
is made, then the rest function can be continued into the entire
complex plain
$t$ and one can accurately examine its properties in the
neighborhood of origin.

It is usually considered that if the function asymptotically tends
to 1
at $t \to 0$ then it is quite enough to fulfil the initial condition.
However, in the case of divergent expansion $t=0$ is essential
singular point, and, really, the function does not have any meaning
at this
point. The function tends to 1 only if its argument belongs to some
sector
in the complex plain. When the argument comes to zero outside this
sector
then one can obtain every desired meaning of the function.

In this case one is to account that, strictly speaking, the evolution
operator kernel does not exist. And these potentials cannot be
considered
in the quantum mechanics as exact ones. They can be treated only as
some
approximations to unknown exact potentials for which the
Schwinger--DeWitt
expansion is convergent. Namely in this context one should consider
the weaken Cauchy's problem, in which one does not impose the rigorous
 demands
to behavior of the solution in the vicinity of the point $t=0$.

In this paper some nontrivial potentials are presented for which the
Schwinger--DeWitt expansion is convergent. Those are the potentials
being
widely used in one-dimensional many-body
problems~\cite{OP,CMR,Suth1,Suth2}.
For definite discrete values of the coupling constant the expansions
for them are convergent in the entire complex plain $t$. For other
values of the charge
the expansions are asymptotic and, so, the kernels in rigorous sense
do not exist. This fact can be treated as quantization of the charge.
The solution
of the Cauchy's problem for the Schr\"odinger equation exists only
for some
discrete values of the charge. Hence, the potentials with those
charges can only be considered in the exact quantum theory.

\section{The method of research}

The evolution operator kernel of the Schr\"odinger equation in
one-dimensional case is the solution of the problem
  \begin{equation} \label{f1}
  \frac{\partial}{\partial t} \langle q',t\mid q,0 \rangle =
  \frac{1}{2} \frac{\partial^2}{\partial q'^2}
  \langle q',t\mid q,0 \rangle - V(q') \langle q',t\mid q,0 \rangle,
  \end{equation}
  \begin{equation} \label{f2}
  \langle q',t=0\mid q,0 \rangle = \delta (q'-q).
  \end{equation}
Here and everywhere below dimensionless values, which are derived from
dimension ones in an obvious way, are used for the sake of
convenience.
The variable $t$ is treated as a complex one. If one means the proper
Schr\"odinger equation then $t$ should be equal to $i\tau$, where
$\tau$ is a real variable (physical time). Because we intend to study
analytical properties of the kernel in variable $t$, then it is
convenient to use $t$
instead of $\tau$. We imply that $V(q)$ does not apparently depend
on time.

As it is well known, in the free case ($V \equiv 0$) the solution of
the problem~(\ref{f1}), (\ref{f2}) is
  \begin{equation} \label{f3}
  \langle q',t\mid q,0 \rangle = \frac{1}{\sqrt{2\pi t}}
  \exp \left\{- \frac{(q'-q)^2}{2t} \right\} \equiv \phi(t;q',q).
  \end{equation}
The function $\phi$ has essential singularity at $t=0$, but this
singularity is such, that it provides the initial condition~(\ref{f2})
 to be fulfilled.

When interaction is present the kernel can be represented as
  \begin{equation} \label{f4}
  \langle q',t\mid q,0 \rangle = \frac{1}{\sqrt{2\pi t}}
  \exp \left\{- \frac{(q'-q)^2}{2t} \right\} F(t;q',q),
  \end{equation}
and besides, one can write for $F$ the expansion (the short-time
Schwinger--DeWitt expansion)
  \begin{equation} \label{f5}
  F(t;q',q) = \sum_{n=0}^{\infty} t^n a_n(q',q),
  \end{equation}
which, as a rule, is asymptotic and is usually utilized only in that
quality. Particularly, in~\cite{S2} it is shown that for the
polynomial potentials (of an
order $L$ higher then 2) expansion~(\ref{f5}) diverges as
$\Gamma \left( \frac{L-2}{L+2}n \right)$.

The question arises if the emphasis of the $\delta$-like function in
form of the
function $\phi$ in~(\ref{f4}) is unique. Generally speaking, every
function of
the form $\Phi((q'-q)/\sqrt{t})/\sqrt{t}$ with normalization
condition
  $$\int _{-\infty}^{+\infty} \!\! \Phi(z)dz=1$$
 will tend to $\delta (q'-q)$ when $t \to 0$. Let us take, in this
connection, the general representation
  \begin{equation} \label{f5'}
  \langle q',t\mid q,0 \rangle = \frac{1}{\sqrt{t}}
  \Phi \left(\frac{q'-q}{\sqrt{t}}\right) \tilde F(t;q',q)
  \end{equation}
and substitute it into~(\ref{f1}). Taking into account, that $\Phi$
depends on the variables only through the combination
$z=(q'-q)/\sqrt{t}$,
but in $\tilde F$ the variables $t,\ q'$, and $q$ vary independently
with each other, one can separate the variables and get
  \begin{equation} \label{f5''}
  \frac{d^2 \Phi}{dz^2} + z \frac{d \Phi}{dz} + \Phi(z)=0,
  \end{equation}
  \begin{equation} \label{f5'''}
  \frac{\partial \tilde F}{\partial t} =
  \frac{1}{2} \frac{\partial ^2 \tilde F}{\partial q'^2} +
  \frac{1}{\sqrt{t} \Phi} \frac{d \Phi}{dz}
  \frac{\partial \tilde F}{\partial q'} - V(q') \tilde F.
  \end{equation}
Demanding $\tilde F$ to be regular at $t=0$ and at $q'=q$ one has
an unambiguous definition of the function $\Phi(z)$
  \begin{equation} \label{f5''''}
  \Phi (z)= \frac{1}{\sqrt{2 \pi}} \exp \{- z^2/2\}.
  \end{equation}
This exactly coincides with~(\ref{f4}) and so, $\tilde F=F$ and
(\ref{f5'''}) with taking into consideration of~(\ref{f5''''}) gives
the equation for the function $F$.

We shall use representation~(\ref{f4}), (\ref{f5}) to test the
analytical properties of the evolution operator kernel in variable $t$
and, in particular, ascertain its behavior for $t \to 0$, which is
necessary to examine whether initial condition~(\ref{f2}) is fulfilled
or not.

For this purpose let us derive from~(\ref{f1}) (or
from~(\ref{f5'''}),~(\ref{f5''''})) the equation for $F$
  \begin{equation} \label{f6}
  \frac{\partial F}{\partial t} =
  \frac{1}{2} \frac{\partial ^2 F}{\partial q'^2} -
  \frac{q'-q}{t} \frac{\partial F}{\partial q'} - V(q')F.
  \end{equation}
Because the factorized function $\phi$ yet fulfils the initial
condition (\ref{f2}), then $F$ should satisfy the initial condition
  \begin{equation} \label{f7}
  F(t=0;q',q)=1.
  \end{equation}
It seems, at first sight, that it is possible to add to the
right-hand side of~(\ref{f7})
an arbitrary function of $q'-q$, which vanishes at $q'=q$. However,
this is not true. The equation for the coefficient $a_0$
  $$(q'-q) \frac{\partial a_0(q',q)}{\partial q'} =0,$$
 taken from general recursion relations for $a_n(q',q)$,
and condition $a_0(q,q)=1$ determines unambiguously
  $$a_0(q',q)=F(0;q',q)=1.$$

The problem~(\ref{f1}), (\ref{f2}), from which we have started, has
a physical
sense only for the real positive $t$ (if the heat equation and heat
kernel are
considered) or for $t=i \tau$, where $\tau$ is the real positive (if
the quantum mechanical evolution equation is considered). The same
restrictions
are initially fair for equation~(\ref{f6}) too. But we can
analytically continue the function $F$ into complex plain of the
variable $t$ using
the differential equation~(\ref{f6}) with condition~(\ref{f7}).

Write the representation of kind~(\ref{f5}) giving concrete form to
the coordinate dependence of the coefficients $a_n$
  \begin{equation} \label{f8}
  F(t;q',q) = 1+
  \sum_{n=1}^{\infty} \sum_{k=0}^{\infty} t^n \Delta q^k b_{nk}(q).
  \end{equation}
Here $\Delta q=q'-q$. For the potential $V$ we write the expansion in
powers of $\Delta q$
  \begin{equation} \label{f9}
  V(q') = \sum_{k=0}^{\infty} \Delta q^k \frac{V^{(k)}(q)}{k!}.
  \end{equation}
The notation similar with
  $$V^{(k)}(q) \equiv \frac{d^k V(q)}{dq^k}$$
will be used further everywhere. We mean that the point $q$
is a regular point of the function $V(q)$ because we are interested
in the regular points only when calculating the kernel.

It is obvious that
  $$\sum_{k=0}^{\infty} \Delta q^k b_{nk}(q)= a_n(q',q)=-Y_n(q',q),$$
where $Y_n$ are the functions introduced in~\cite{S1}. The behavior
of $Y_n$ was studied in~\cite{S2} using representation adduced
in~\cite{S1}.

Substitution of~(\ref{f8}), (\ref{f9}) into~(\ref{f6}) leads to
recurrent relations for the coefficients $b_{nk}$
  \begin{equation} \label{f10}
  b_{nk}=\frac{1}{n+k} \left[\frac{(k+1)(k+2)}{2} b_{n-1, k+2} -
  \sum_{m=0}^k \frac{V^{(m)}(q)}{m!} b_{n-1, k-m} \right]
  \end{equation}
with condition $b_{0k}= \delta_{k0}$. Specifically,
  \begin{equation} \label{f11}
  b_{1k}=- \frac{V^{(k)}(q)}{(k+1)!}.
  \end{equation}

Expressions~(\ref{f8}), (\ref{f10}) determine a formal solution of
problem~(\ref{f6}),~(\ref{f7}). As to the expansion in powers of
$\Delta q$ in~(\ref{f8}), one may expect that its convergence range
is equal to the one for expansion~(\ref{f9}) of the potential. The
series in $t$
in~(\ref{f8}) is usually treated as the divergent one. At first sight,
 it is really so always. Let us estimate the convergence of the
 series in~(\ref{f8}).

At the beginning let $n$ be fixed and $k \to \infty$. Expressing
$b_{n-1, k+2}$
from~(\ref{f10}) via the coefficients with smaller $k$ we shall come
to some linear combination of the coefficients of type $b_{n_0, 0}$
and $b_{n_1, 1}$ with any indexes $n_0, n_1$ (for the sake of brevity
we shall
write further only the terms with $b_{n_0, 0}$, implying that the same
statements are concerned with the terms with $b_{n_1, 1}$). The main
growth for
large $k$ takes place if the second index of $b_{nk}$ is diminished:
a)~with using the term $V^{(k)} b_{n-1, 0}/k!$, b)~with using
the expression on the left-hand side of~(\ref{f10}).

In the case a) we get for $k \to \infty$
  \begin{equation} \label{f12}
  |b_{n-1, k+2}^{(a)}| \sim \frac{2}{(k+1)(k+2)} \frac{|V^{(k)}|}{k!}
  |b_{n-1, 0}|.
  \end{equation}
Because series~(\ref{f9}) converges at some circle with
radius $R(q)$ the estimate
  $$\frac{|V^{(k)}|}{k!} \sim \frac{1}{R^k(q)}$$
for $k \to \infty$ is fair. So, for every fixed $n$ and for
$k \to \infty$ we have
  \begin{equation} \label{f13}
  |b_{nk}^{(a)}| \sim \frac{|b_{n0}|}{R^k(q)}.
  \end{equation}
The contributions of type~(\ref{f13}) correspond to the expansion
in $\Delta q$,
which is convergent for every fixed $n$ with convergence
range $R(q)$.

In the case b) for $k \to \infty$ we get
  $$ |b_{n-1, k+2}^{(b)}| \sim \frac{2^{k/2+1} (n+k)!}{k! (n+k/2-1)!}
     |b_{n+k/2, 0}|.$$
Behavior of $b_{nk}^{(b)}$ for $k \to \infty$ depends on the behavior
 of $b_{n0}$ for $n \to \infty$. If $b_{n0}$ decreases when
 $n \to \infty$ or
increases more slowly than $\Gamma(\alpha n)$ ($\alpha$ is any
positive number), then
  $$|b_{nk}^{(b)}| \sim \frac{|b_{n0}|}{\Gamma (k/2)}$$
for $k \to \infty$, i.e., these contributions will disappear at
large $k$.
If $b_{n0}$ increases as $\Gamma(\alpha n)$ (here $0< \alpha \le 1$,
in~\cite{S2} showed that $\alpha$ cannot be larger then 1), then for
$k \to \infty$ and $\alpha < 1$
  $$|b_{nk}^{(b)}| \sim
       \frac{|b_{n0}|}{\Gamma \left( \frac{1-\alpha}{2}k \right)},$$
so, these contributions will disappear too with the
growth of $k$. If $\alpha =1$,
then the following estimate will take place ($n$ is fixed,
$k \to \infty$)
  \begin{equation} \label{f15}
  |b_{nk}^{(b)}| \sim |b_{n0}| k^c \rho ^k.
  \end{equation}
In this case the expansion in $\Delta q$ in~\ref{f8}) will have the
finite convergence range too, but it will be equal to minimum from
two values $R(q)$ and $\rho$.

Now let us examine the behavior of $|b_{n0}|$ (the same will be also
correct for
$|b_{n1}|$) when $n \to \infty$. Consider the decreasing of $n$ till
1 by means of the first term on the right-hand side of~(\ref{f10})
  \begin{equation} \label{f16}
  |b_{n0}| \sim \frac{|b_{n-1, 2}|}{n} \sim \cdots \sim
  \frac{(n-1)!}{2^{n-1}} |b_{1,2n-2}|=
  \frac{(n-1)!}{2^{n-1}} \frac{|V^{(2n-2)}|}{(2n-1)!}.
  \end{equation}
Because $|V^{(k)}| \sim k!/R^k(q)$ for $k \to \infty$, then
for $n \to \infty$ we get
  \begin{equation} \label{f17}
  |b_{n0}| \sim \frac{(n-1)!}{2^{n-1} (2n-1)} \sim n!.
  \end{equation}

Really, the contributions taken into account in~(\ref{f16})
provide the main growth only for the potentials, for which
$R(q)<\infty$. If the potentials with $R(q)=\infty$ are
considered (e.g., polynomial ones), then, at first sight, one can
conclude
from~(\ref{f16}) that $|b_{n0}| \sim 1/n!$. But it is not so, in fact.
As it was shown in~\cite{S2}, the combination of contributions of the
 first
term and terms of sum over $m$ in~(\ref{f10}) leads to the estimate of
 type $|b_{n0}| \sim \Gamma(\alpha n)$.

So, for arbitrary potentials the series in $t$ in~(\ref{f8}) is
divergent.
But in our estimates, in fact, absolute values of all contributions to
every coefficient $b_{nk}$ were summed. Nevertheless, for some
potentials
the cancellation of different terms may occur. It can lead to
convergence of
the expansion in~(\ref{f8}). For the potentials considered in
Secs.~3--5 this
cancellation takes place only for definite values of the coupling
constant.

Note that we, really, test expansion~(\ref{f8}) for the absolute
convergence.
So, it is enough for the convergence of double series that~(\ref{f8}),
in which
instead of $b_{nk}$ absolute values $|b_{nk}|$ taken, would converge
for any
order of summation. Our consideration corresponds to the following
order: at
first the series over $k$ for every fixed n are summed and then
summation
over $n$ is made. If one assumes that there is convergence of the
series
in index $n$ then, as it was shown before, the convergence in index
$k$ will
take place at every fixed $n$, and to establish the convergence of
the series
in index $n$ it is enough to determine the behavior  of the
coefficients $b_{n0}, \ b_{n1}$ only (but not all $b_{nk}$) at
$n \to \infty$.

\section{Potential $V(q)=g/q^2$}

Let us introduce standard notation for the coupling constant
$g=\lambda
(\lambda -1)/2$ ($\lambda >0$) and investigate the potential
  \begin{equation} \label{f18}
  V(q)= \frac{\lambda (\lambda -1)}{2} \frac{1}{q^2}
  \end{equation}
for the convergence of expansion~(\ref{f8}).

Expansion~(\ref{f9}) for potential~(\ref{f18}) has the finite
convergence
range $R(q)=q$ which is connected with singularity of $V(q)$ at
the point $q=0$. The derivatives $V^{(k)}$ may be easily calculated
  \begin{equation} \label{f19}
  V^{(k)}(q)= (-1)^k \frac{\lambda (\lambda -1)}{2}
  \frac{(k+1)!}{q^{k+2}}.
  \end{equation}
{}From~(\ref{f10}) with account of~(\ref{f19}) we take
  \begin{equation} \label{f20}
  b_{nk}=\frac{1}{n+k} \left[\frac{(k+1)(k+2)}{2} b_{n-1, k+2} +
  \frac{\lambda (\lambda -1)}{2} \sum_{m=0}^k (-1)^{m+1}
  \frac{m+1}{q^{m+2}} b_{n-1, k-m} \right].
  \end{equation}
If we shall diminish $n$ times the first index of $b_{nk}$ by means
of~(\ref{f20}), then we get
  \begin{equation} \label{f21}
  b_{nk}= \frac{(-1)^{n+k}}{q^{2n+k}} \frac{(k+n-1)!}{n!(n-1)!k!}
        \prod_{j=1}^n \left(\frac{\lambda (\lambda -1)}{2}
        - \frac{j(j-1)}{2} \right).
  \end{equation}

For noninteger $\lambda$ the main growth at $n \to \infty$ of the
polynomial of the order $n$ in
the variable $g=\lambda (\lambda -1)/2$  which we denote as
  \begin{equation} \label{f22}
  \Lambda_n(g)= \prod_{j=1}^n \left( g- \frac{j(j-1)}{2} \right)
  \end{equation}
will be provided by the coefficient in front of the first power
of $g$. This coefficient is equal to $n!(n-1)!/2^{n-1}$. So, for
$n \to \infty$ the estimate
  \begin{equation} \label{f23}
  |b_{nk}| \sim \frac{(n+k-1)!}{2^{n-1} |q|^{2n+k}k!} \sim n!
  \end{equation}
is true. It means that for noninteger $\lambda$ expansion~(\ref{f8})
for potential~(\ref{f18}) is divergent.

Consider now integer $\lambda$ ($\lambda >1$, because cases
$\lambda =0, \ \lambda =1$ are trivial). In this case $\Lambda_n =0$
for $n \ge \lambda$. Hence, only the coefficients $b_{nk}$ for
$n< \lambda$ are
different from zero, and in~(\ref{f8}) the series in powers of $t$ is
really the polynomial of finite degree $\lambda -1$.

Let us substitute~(\ref{f21}) (for $n< \lambda$) into~(\ref{f8}) and
make summation over $k$. Then we get finally
  \begin{equation} \label{f24}
  F(t;q',q)= 1+ \sum_{n=1}^{\lambda -1} \frac{(-1)^n t^n}{q'^n q^n}
  \frac{1}{n!} \prod_{j=1}^n \left(\frac{\lambda (\lambda -1)}{2}
        - \frac{j(j-1)}{2} \right).
  \end{equation}

Rigorously speaking, the derivation of~(\ref{f24}) is made in
supposition
that $|\Delta q| < |q|$. If this condition is not satisfied, then
calculations should be made with the expansion about the point
$q'$ in powers of $q-q'$. But because
$F$ is symmetric in $q', \ q$, then it is clear that the answer
in this case would be the same as in~(\ref{f24}). These reasonings
are valid for the case when $q$ and $q'$ have the same sign. But the
potential~(\ref{f18}) at $q \to 0$ tends to $\infty$, and the
transition through
the point $q=0$ is impossible. Hence, the kernel is equal to zero if
$q$ and $q'$ have different signs.

So, we established that for integer $\lambda$ expansion~(\ref{f4}),
(\ref{f8}) for the evolution operator kernel is not asymptotic, but
it is the convergent one. The function $F$ is presented by the
polynomial of degree
$\lambda -1$ in powers of $t$ and inverse powers of $q'$ and $q$.
It is single-valued and analytic in the entire complex plain of the
variable $t$ function. There is the pole of the
order $\lambda -1$ in the infinite point. Further consideration
for noninteger $\lambda$ will be made in Sec.~6.

\section{Modified P\"oschl--Teller's potential $V(q)=-g/\cosh ^2q$}

Another example of convergent series~(\ref{f8}) we shall get
considering the potential
  \begin{equation} \label{f25}
  V(q)= - \frac{\lambda (\lambda -1)}{2} \frac{\beta^2}
  {\cosh ^2(\beta q)}.
  \end{equation}
Because the constant $\beta$ is connected with the choice of
length scale one
can put $\beta=1$ without the restriction of generality.
Further, for the sake of brevity we shall denote
  \begin{equation} \label{f26}
  f(q)= - \frac{1}{\cosh ^2q}.
  \end{equation}
Then the potential reads briefly $V(q)=gf(q)$.

The potential~(\ref{f25}) has the expansion of type~(\ref{f9}) about
every real  point $q$. Its convergence range is equal to $R(q)=
\sqrt{(\pi/2)^2+q^2}$
and is determined by the distance to the nearest singularities of the
function $1/\cosh ^2 q$ placed at the points $q= \pm i \pi/2$.
 The derivatives can be calculated as follows
  \begin{equation} \label{f27}
  V^{(k)}(q)= gf^{(k)}(q),
  \end{equation}
where $f^{(k)}$ are represented as expansions in powers of $f$
  \begin{equation} \label{f28}
  f^{(2n)}(q)= \sum_{l=1}^{n+1} a_l^{(2n)} f^l(q),
  \end{equation}
  \begin{equation} \label{f29}
  f^{(2n+1)}(q)= \sum_{l=1}^{n+1} la_l^{(2n)} f^{l-1}f^{(1)}=
                 \sum_{l=1}^{n+1} a_l^{(2n+1)} f^{l-1}f^{(1)}.
  \end{equation}

To obtain all coefficients $a_l^{(k)}$ it is enough to put $a_l^{(0)}=
\delta_{l1}$ and take into account
  $$(f^{(1)})^2=4f^3+4f^2.$$
For $a_l^{(2n)}$ one has the recursion relations
  \begin{equation} \label{f30}
  a_l^{(2n)}= 4l^2 a_l^{(2n-2)} +4(l-1)(l-1/2) a_{l-1}^{(2n-2)}.
  \end{equation}
So, every derivative of the function $f(q)$ is represented as a
polynomial in powers of this function.

{}From~(\ref{f10}) one gets for potential~(\ref{f25})
  \begin{equation} \label{f31}
  b_{nk}=\frac{1}{n+k} \left[\frac{(k+1)(k+2)}{2} b_{n-1, k+2} -
  \frac{\lambda (\lambda -1)}{2} \sum_{m=0}^k
  \frac{f^{(m)}}{m!} b_{n-1, k-m} \right],
  \end{equation}
where the derivatives $f^{(m)}$ are calculated
via~(\ref{f28})--(\ref{f30}).

According to the note at the end of Sec.~2, it is enough
for testing the convergence of series~(\ref{f8})
to examine the behavior at $n \to \infty$ of the coefficients
$b_{n0}, \ b_{n1}$ only. Introduce in this connection the functions
  \begin{equation} \label{f32}
  B_k(t,q)= \sum_{n=0}^{\infty} t^n b_{nk}(q)
  \end{equation}
and consider them for $k=0,\ 1$.

The analysis of relations~(\ref{f31}) with taking into account
of~(\ref{f28})--(\ref{f30}) shows that $B_0, \ B_1$ can be
represented in the form
  \begin{equation} \label{f33}
  B_0(t,q)=1+ \sum_{n=1}^{\infty} t^n \sum_{l=1}^n
           \frac{(-1)^l}{l!} f^l(q) \beta_{nl}
           \prod_{j=1}^l \left(\frac{\lambda (\lambda -1)}{2}
           - \frac{j(j-1)}{2} \right),
  \end{equation}
  \begin{equation} \label{f34}
  B_1(t,q)= \sum_{n=1}^{\infty} t^n \sum_{l=1}^n
           \frac{(-1)^l}{l!} \frac{l}{2} f^{l-1}(q) f^{(1)}(q)
           \beta_{nl}
           \prod_{j=1}^l \left(\frac{\lambda (\lambda -1)}{2}
           - \frac{j(j-1)}{2} \right),
  \end{equation}
where
  \begin{equation} \label{f35}
  \beta_{nl}= \frac{1}{2^{n-l}} \frac{(n-1)!}{(l-1)!}
                 \frac{a_l^{(2n-2)}}{(2n-1)!}.
  \end{equation}

To estimate the behavior of $\beta_{nl}$ when $n \to \infty$ we
probe the asymptotics of $a_l^{(2n-2)}$. Let us take $a_l^{(2n-2)}$
for sufficiently
large $n$ and begin to express $a_l^{(2n-2)}$ with the help
of~(\ref{f30}) via
the coefficients with smaller $n$ and $l$ so that to come to
$a_1^{(0)}=1$
at the end. Maximal contribution arises in this procedure when,
at the beginning, $n$ will be diminished at fixed $l$ by means of
the first term
on the right-hand side of~(\ref{f30}), and then, when the
equality $n=l-1$
becomes valid, $n$ and $l$ will start to be decreased simultaneously
by unit at
every step by means of the second term in~(\ref{f30}). For large $n$
this gives the estimate
  $$a_l^{(2n-2)} \sim 4^{n-l} l^{2(n-l)} (2l-1)!.$$
Then $\beta_{nl}$ behaves itself as
  \begin{equation} \label{f37}
  \beta_{nl} \sim 2^{n-l} l^{2(n-l)} \frac{(n-1)!}{(l-1)!}
                 \frac{(2l-1)!}{(2n-1)!}.
  \end{equation}

Now one can evaluate the asymptotics at $n \to \infty$ of the
coefficients of series~(\ref{f33}), (\ref{f34}). For noninteger
$\lambda$ at
sufficiently large $l$ the polynomial $\Lambda_l(g)$ (see~(\ref{f22}))
behaves itself as $\Lambda_l \sim g l!(l-1)!/2^{l-1}$. Taking
in~(\ref{f33}), (\ref{f34}) in the sum over $l$ the term with $l=n$
we shall obtain that the coefficients in front of $t^n$ growth
in~(\ref{f33}) as
  $$\frac{(n-1)!f^n}{2^{n-1}},$$
and in~(\ref{f34}) as
  $$\frac{n!f^{n-1}f{(1)}}{2^n}.$$
So, for noninteger $\lambda$ series~(\ref{f33}),~(\ref{f34}),
and, hence, (\ref{f8}) are asymptotic ones.

Let now $\lambda$ be integer ($\lambda >1$). Then the polynomial
$\Lambda_l(g)$ is different from zero only if $l< \lambda$. So,
in~(\ref{f33}),~(\ref{f34}) in the sums over $l$ only the terms with
$l< \lambda$ are different from zero, and, in fact, one should take
instead of
$\sum_{l=1}^n$ the sum $\sum_{l=1}^{\min \{n,\lambda-1\} }$.
For $n \ge
\lambda -1$ the sum over $l$ will always contain the same number of
terms,
equal to $\lambda -1$, and its dependence on $n$ will be determined
only
by the dependence on $n$ of the coefficients $\beta_{nl}$.
And the dependence of the latter on $n$,
as it is clear from estimate~(\ref{f37}), at fixed
$l \le \lambda -1$ and
at $n \to \infty$ is determined by the factor
  $$\beta_{nl} \sim \left(2(\lambda -1)^2 \right)^n
  \frac{(n-1)!}{(2n-1)!}.$$
So, the coefficients in front of $t^n$ in~(\ref{f33}),~(\ref{f34})
behave themselves at large $n$ as
  $$\frac{C^n (n-1)!}{(2n-1)!},$$
i.e., the series will be convergent at the circle of infinite range.

To obtain finally the function $F(t;q',q)$ it is necessary either to
 take the
coefficients $b_{n0}, \ b_{n1}$ from~(\ref{f33}),~(\ref{f34}) to
calculate
other $b_{nk}$ using~(\ref{f31}), or starting from $B_0, \ B_1$ to
calculate
other functions $B_k(t,q)$ from the equations
  \begin{equation} \label{f38}
  B_{k+2}=\frac{2}{(k+1)(k+2)} \left( \frac{\partial B_k}
  {\partial t} +
    \frac{k}{t} B_k + \sum_{m=0}^k \frac{f^{(m)}}{m!}
    B_{k-m} \right),
  \end{equation}
which in an obvious way are derived from~(\ref{f6}) after
substitution
  \begin{equation} \label{f39}
  F(t;q',q)= \sum_{k=0}^{\infty} \Delta q^k B_k(t,q),
  \end{equation}
and substitute them into~(\ref{f39}).

Particularly, for $\lambda =2 \ (g=1)$ we have the potential $V(q)=
-1/\cosh^2 q$, for which
  \begin{equation} \label{f40}
  B_0(t,q)=1- f(q) \sum_{n=1}^{\infty} \frac{t^n}{(2n-1)!!}=
           1- f(q) \sqrt{\frac{\pi t}{2}} e^{t/2} \erf (\sqrt{t/2}),
  \end{equation}
  \begin{equation} \label{f41}
  B_1(t,q)=-\frac{1}{2} f^{(1)}(q) \sum_{n=1}^{\infty}
  \frac{t^n}{(2n-1)!!}=
   -\frac{1}{2} f^{(1)}(q) \sqrt{\frac{\pi t}{2}} e^{t/2}
   \erf (\sqrt{t/2}).
  \end{equation}
With the help of~(\ref{f38}) one is able to determine all
coefficient functions
$B_k$ starting from~(\ref{f40}),~(\ref{f41}) and then to substitute
them into~(\ref{f39}). In this manner the function $F$ will be found.

We established, that for integer $\lambda$ expansion~(\ref{f8}) was
convergent if $|\Delta q| < R(q)$ and the representation~(\ref{f4}),
{}~(\ref{f8}) for the evolution operator kernel was not asymptotic. The
 function
$F$ is single-valued analytic in the entire complex plain of the
variable $t$ function and it has an essential singularity at the
infinite $(t=\infty)$ point.

\section{Other samples of potentials}

Calculations made in Secs.~3,~4 may be easily repeated for some
similar potentials which are often used in one-dimensional many-body
problems~\cite{OP,CMR,Suth1,Suth2}.

\subsection{Potential $V(q)=g/ \sinh^2 q$}

Analogously to~(\ref{f25}) one can consider the potential
  \begin{equation} \label{f42}
  V(q)= \frac{\lambda (\lambda -1)}{2} \frac{1}{\sinh ^2q}.
  \end{equation}
Denote
  \begin{equation} \label{f43}
  f(q)= \frac{1}{\sinh ^2q},
  \end{equation}
and notice, that
  $$(f^{(1)})^2=4f^3+4f^2,$$
i.e., it exactly coincides with the
corresponding expression for the function $f(q)$ defined
by~(\ref{f26}) in
Sec.~4. So, all relations for the derivatives of $f$ obtained
there and,
hence, expressions for $b_{nk}, \ B_k$, and $F$ remain right in
this case.
There exists only one difference: function $f$ is defined now
by~(\ref{f43}), but not by~(\ref{f26}).

One can conclude that the kernel for potential~(\ref{f42}) for
integer $\lambda$ exists and is determined by equations~(\ref{f4}),
(\ref{f33})--(\ref{f35}), (\ref{f38}), (\ref{f39}), (\ref{f43}). The
function $F$ is single-valued and analytic function in the entire
complex plain of the variable $t$.

\subsection{Potential $V(q)=g/ \sin^2 q$}

The potential
  \begin{equation} \label{f44}
  V(q)= \frac{\lambda (\lambda -1)}{2} \frac{1}{\sin ^2q}
  \end{equation}
can be also considered in similar way. Denote
  \begin{equation} \label{f45}
  f(q)= \frac{1}{\sin ^2q}
  \end{equation}
and take into account, that
  $$(f^{(1)})^2=4f^3-4f^2.$$
This expression
differs from analogous ones for potentials~(\ref{f25}),~(\ref{f42})
only by the sign of the second term. So, we are able to reconstruct
the expressions from Sec.~4 with small changes only:
in~(\ref{f33}),~(\ref{f34}) will appear an additional multiplier
$(-1)^{n+l}$, and the function $f$ will be defined by~(\ref{f45}).
The conclusion about the existence of the kernel for integer
$\lambda$ remains right for the potential~(\ref{f45}).

\section{Divergent expansions. Quantization of charge}

Now we shall consider the potentials discussed above, but take
noninteger $\lambda$. For this we, at first, establish the character
of singularity
of the function $F$ at $t=0$, which causes the divergence of
expansion~(\ref{f8}). In general case the following variants are
possible:
1)~$F$ is the single-valued function of $t$ and the point $t=0$
is the singular
point of one-valued character; 2)~$F$ is the many-valued function
and the point $t=0$ is the branching point of finite order $N-1$;
3)~$F$ is the many-valued
function and the point $t=0$ is the logarithmic branching point.

Show, at first, that the case 3) is really excluded, i.e., that
the solution
of~(\ref{f6}) cannot have the logarithmic branching point at $t=0$.
 Assume the contrary statement: the point $t=0$ is the logarithmic
branching point of $F$,
where $F$ is the solution of~(\ref{f6}). Then in any neighborhood
of zero $G=\{t: 0<|t|<\rho\}$ the representation
  \begin{equation} \label{f45'}
  F(t;q',q)= \sum_{n=0}^{\infty}  \left( \frac{\Ln t -\alpha}
             {\Ln t +\alpha}\right)^n A_n(q',q)
  \end{equation}
takes place. Here $\alpha$ is some number, satisfying, in general
case, the
condition $\Re \alpha < \ln \rho; \ \Ln t$ means the full many-valued
logarithmic function. Note, that~(\ref{f45'}) is the simplified
version of
general representation, in which $\alpha$ is taken real and $\rho =1$.

Let us introduce the new variable
  \begin{equation} \label{f46}
  w= \frac{\Ln t -\alpha}{\Ln t +\alpha}.
  \end{equation}
This transformation transfers the region $G$ into the unit circle in
the complex plain $w$. The point $w=1$ corresponds to $t=0$.
Expansion~(\ref{f45'}) takes the form
  \begin{equation} \label{f47}
  F(w;q',q)= \sum_{n=0}^{\infty} w^n A_n(q',q),
  \end{equation}
and according to the assumption, should be convergent in the unit
circle. We
substitute~(\ref{f47})  into~(\ref{f6}) taking into
account~(\ref{f46})
and expand all arising expressions in powers of $w$. This leads
to the equation
  \begin{eqnarray} \label{f48}
  &&\frac{1}{2\alpha} \sum_{n=1}^{\infty} nA_n \left[
    \left( 1-\alpha +\frac{\alpha^2}{2} \right) w^{n-1}
  - (2-\alpha^2)w^n+\left( 1+\alpha +\frac{\alpha^2}{2}\right)
  w^{n+1} \right.
  \nonumber \\*
  &&\left. + \sum_{k=0}^{\infty} \sum_{l=1}^{\infty} \sum_{m=0}^{2l+2}
  \frac{(-\alpha)^{l+2}}{(l+2)!} {{2l+2}\choose m} w^{2lk+m+n-1}
   \right]
  \nonumber \\*
  &&=\sum_{n=0}^{\infty} \left( \frac{1}{2}
     \frac{\partial^2 A_n}{\partial q'^2}
  - V(q')A_n - \Delta q \frac{\partial A_n}{\partial q'} \right) w^n
  \nonumber \\*
  &&- \sum_{n=0}^{\infty} \sum_{k=0}^{\infty} \sum_{l=1}^{\infty}
  \sum_{m=0}^{2l} \Delta q \frac{\partial A_n}{\partial q'}
  \frac{(-\alpha)^l}{l!} {{2l}\choose m} w^{2lk+m+n}.
  \end{eqnarray}

After equating the coefficients in front of the same powers of $w$
the function $A_{n+1}$ can be expressed via $A_i$ with $i \le n$ and
their derivatives with respect to $q'$. For example,
  \begin{equation} \label{f49}
  A_1= 2\alpha e^{\alpha} \left( \frac{1}{2}
      \frac{\partial^2 A_0}{\partial q'^2}  - V(q')A_0 -
      e^{-\alpha} \Delta q \frac{\partial A_0}{\partial q'} \right).
  \end{equation}

Let us take sufficiently large $n$ and calculate the coefficients
$A_{n+1}$
through $A_i$. We shall not write the relations between $A_i$ in
the general
form, but write only the terms, which will be considered.
{}From~(\ref{f48}) we get
  \begin{equation} \label{f50}
  \frac{e^{-\alpha}}{2\alpha} \left\{ (n+1)A_{n+1} - 2(1+\alpha)nA_n
  \right\}= \{ \mbox{{\em other terms depending on }}
  A_i, \ i \le n \}.
  \end{equation}
The contributions into $A_{n+1}$, arising at the transition
$A_{i+1} \to A_i$ by means of the terms displayed in~(\ref{f50}),
will be considered now
  \begin{equation} \label{f51}
  |A_{n+1}| \sim \frac{2|1+\alpha |n}{n+1} |A_n| \sim \cdots \sim
           \frac{2^n |1+\alpha |^n}{n+1} |A_0|.
  \end{equation}
The existence of these contributions means that expansion~(\ref{f47})
 can converge at the circle of radius not larger then
  $$R_w=\frac{1}{2|1+\alpha|}.$$
Because $\alpha$ is an arbitrary negative number, it is clear that
$R_w$ may be done by choosing the corresponding $\alpha$ as small
as one whishes.
But if the point $t=0$ were the logarithmic branching point of $F$
then expansion~(\ref{f47}) would be convergent at the circle of unit
 radius.
We have a contradiction, so the solution of~(\ref{f6}) cannot have
the logarithmic branching point at $t=0$.

Let us consider now case 1). According to known theorem~\cite{Olv},
 the point $t=0$, if the series~(\ref{f8}) (or~(\ref{f5})) is
 divergent, is
an essential singular point of the function $F$. And what's more,
expansion~(\ref{f8}) approximates  $F$ only in some sector of
the complex plain $t$ with angle $\gamma <2 \pi$. The function $F$
does not have any meaning at the point $t=0$.

Case 2). The point $t=0$ is the branching point of $F$ of
the order $N-1$. Then
in some neighborhood of zero it can be represented by the
convergent series
  \begin{equation} \label{f52}
  F(t;q',q) = \sum_{n=-\infty}^{+\infty} t^{n/N} d_n(q',q).
  \end{equation}

Now we exclude the variant, when the function $F$ has any definite
 meaning at $t=0$ and, at the same time, is represented by divergent
series~(\ref{f8}). Suppose that instead of~(\ref{f52}) the expansion
  \begin{equation} \label{f53}
  F(t;q',q) = \sum_{n=0}^{\infty} t^{n/N} d_n(q',q)
  \end{equation}
takes place (the function having the branching point at zero will
have definite meaning at this point only if~(\ref{f53}) is
convergent).
Substitution of~(\ref{f53}) into~(\ref{f6}) will give the
recurrent relations
for $d_n$. There will be $N$ independent series of the coefficients
$d_n$ having the following forms: $d_{Nl}, \ d_{Nl+1}, \dots,
d_{Nl+N-1}, \ l=0,\, 1, \, 2, \dots$ The coefficients of every
series are expressed
through each other only, but not through the coefficients of other
series.

Let us take the series with numbers of the form $n=Nl$. The equations
 for its coefficients read
  \begin{equation} \label{f54}
  ld_{Nl}+(q'-q) \frac{\partial d_{Nl}}{\partial q'} =
  \frac{1}{2} \frac{\partial^2 d_{N(l-1)}}
  {\partial q'^2} -V(q')d_{N(l-1)}.
  \end{equation}
If one compares $d_{Nl}(q',q)$ with $a_l(q',q)$, where $a_l$ are the
coefficients from~(\ref{f5}), then one can see that the equations for
$d_{Nl}$ are the same as the ones for $a_l$. So, if
expansion~(\ref{f5}) is
divergent, then~(\ref{f53}) is divergent too.

We can state now, that the divergence of series~(\ref{f5}),~(\ref{f8})
means
that $F$ may be either single-valued or $N$-valued with the branching
point
at origin, but in both cases the point $t=0$ is the essential singular
point
(to be exact, in many-valued case $F$ is the single-valued function
of the variable $t^{1/N}$ and the point $t^{1/N}=0$ is the essential
 singular point of
this  function). So, the function $F$ does not have any definite
meaning at $t=0$ when series~(\ref{f5}),~(\ref{f8}) are divergent.

Because  the asymptotic representation~(\ref{f5}) for $F$ is correct
at some sector of the complex plain $t$,
 then if one approaches zero
in bounds of this sector the function $F$ tends to 1. But it does not
mean, that $F(t=0)=1$, because if one approaches zero in other ways,
lying beyond this sector, one can obtain every initially chosen
meaning.

Note, that in~\cite{Mil} the solutions of the complex Schr\"odinger
equation
for the potential~(\ref{f18}) are presented. And they have in general
case the form of the Loran series in powers of $t$.

As a rule, solving the Schr\"odinger equation~(\ref{f1}), one
ignores the fact, that for a number of potentials the solution has
more strong singularity at $t=0$,
than $\delta (q'-q)$, and so, the initial condition~(\ref{f2}) for
these solutions is not satisfied. Practically, it means, that the
solution of
problem~(\ref{f1}),~(\ref{f2}) does not exist, although the solution
of equation~(\ref{f1})~ exists. So, the evolution operator kernel,
strictly speaking, does not exist for the potentials, for which
expansion~(\ref{f5}) is asymptotic. Hence, if we wish to remain in the
framework of conventional formalism of the quantum mechanics, then we
should take into account that these potentials are not
acceptable in the quantum theory.

It is noteworthy, that such strong statement is true only with
respect to
the exact theory. If one considers some model which is meant initially
only as approximate, then there are no reasons to demand that Cauchy's
problem would have the solution in the rigorous sense.
One can consider for that
model potential the weaken Cauchy's problem, in which the strict
requirements
are not imposed on the behavior of the solution at $t \to 0$.
Nevertheless, one should understand that this theory can be treated
 only as an
approximation according to other, usually unknown, exact theory, for
which the evolution operator kernel exists in the rigorous sense,
and one should be ready to operate with asymptotic expansions.

{}From the physical viewpoint the fact of existence of the kernel only
for discrete values of the coupling constant one can treat as natural
quantization of the charge in the quantum theory. Thus, the
potentials of
kind~(\ref{f18}),~(\ref{f25}),~(\ref{f42}),~(\ref{f44}) can be
considered
in the quantum mechanics as exact ones only if at the coupling
constant
$g=\lambda (\lambda -1)/2$ the number $\lambda$ is integer. Of
course, these potentials are the model ones and they have not direct
relation to
quantization of, e.g., electric charge. Nevertheless, the mechanism of
appearance of charge discreteness from the requirement that the
solution of Cauchy's problem for the quantum equation of motion exists
can work in application to more realistic models.

At first sight, the statement about unacceptability in the quantum
theory
the potentials with arbitrary coupling constants (for which the
Hamiltonian
is self-adjoint) seems unexpected. Nevertheless, if one takes into
account,
that the charge in the nature is discrete and, on the other hand,
that the
charge is a parameter of the potential, then it becomes clear that
a correct theory must be constructed so as only the potentials with
quantized values of the coupling constants may be acceptable.

Thus, thorough analysis of analytical structure of the solutions of
the Schr\"odinger equation allows us to discover the potentials, for
which the solution of the rigorous Cauchy's problem exists and which
may be
considered as exact ones. For other potentials only weaken Cauchy's
problem may be formulated and these potentials may be used only as
approximate ones. Moreover, it occurs that for the potentials
considered the kernel exists not for all values of the coupling
constant, but only for some discrete values, i.e., quantization
of the charge takes place.

\section*{Acknowledgments}

It is a pleasure to acknowledge N.~E.~Tyurin for support of my work,
Theory Division of IHEP, where this research has been completed, for
hospitality, and
A.~A.~Arkhipov, V.~A.~Petrov, A.~P.~Samokhin, P.~A.~Saponov,
S.~N.~Storchak and V.~M.~Zhuravlev for useful discussions.

\end{document}